\documentclass[pra,aps,groupaddress,onecolumn]{revtex4}
\usepackage{epsfig,amsmath,graphicx,amssymb,overpic}
\usepackage{bm}
\usepackage{graphicx}
\usepackage{subfigure}
\usepackage{bm}
\usepackage{graphics}
\usepackage{amsfonts}
\usepackage{caption}
\usepackage{epsfig}
\usepackage{epstopdf}
\usepackage{float}
\usepackage{color}

\def\be{\begin{equation}}
\def\ee{\end{equation}}
\def\bee{\begin{eqnarray}}
\def\ene{\end{eqnarray}}
\def\bes{\begin{subequations}}
\def\ees{\end{subequations}}

\newcommand{\sech}{{\rm sech}}

\newcommand{\PT}{\mathcal{PT}}

\begin{document}

\title{Impact of Near-$\PT$ Symmetry on Exciting Solitons and Interactions \\ Based on a Complex Ginzburg-Landau Model}

\author{Yong Chen$^{1,2}$}
\author{Zhenya Yan$^{1,2}$
\footnote{Corresponding author.\, {\it Email address}: zyyan@mmrc.iss.ac.cn (Z. Yan).}}
\author{Wenjun Liu$^{3}$}
\affiliation{\vspace{0.1in}
$^1$Key Lab of Mathematics Mechanization, Academy of Mathematics and Systems Science,  Chinese Academy of Sciences, Beijing 100190, China\\
$^2$School of Mathematical Sciences, University of Chinese Academy of Sciences, Beijing 100049, China \\
$^3$State Key Laboratory of Information Photonics and Optical Communications, School of Science, and P. O. Box 91, Beijing University of Posts and Telecommunications, Beijing 100876, China}

\baselineskip=15pt

\begin{abstract} \baselineskip=15pt

  {\bf Abstract.} \, We present and theoretically report the influence of a class of near-parity-time-($\PT$-) symmetric potentials with spectral filtering parameter $\alpha_2$ and nonlinear gain-loss coefficient $\beta_2$ on solitons in the complex Ginzburg-Landau (CGL) equation. The potentials do not admit entirely-real linear spectra any more due to the existence of coefficients $\alpha_2$ or $\beta_2$. However, we find that most stable exact solitons can exist in the second quadrant of the $(\alpha_2, \beta_2)$ space, including on the corresponding axes. More intriguingly, the centrosymmetric two points in the $(\alpha_2, \beta_2)$ space possess imaginary-axis
 (longitudinal-axis) symmetric linear-stability spectra. Furthermore, an unstable nonlinear mode can be excited to another stable nonlinear mode by
 the adiabatic change of $\alpha_2$ and $\beta_2$. Other fascinating properties associated with the exact solitons are also examined
 in detail, such as the interactions and energy flux. These results are useful for the related experimental designs and applications.


\end{abstract}

\maketitle

\section{Introduction}

The cubic complex Ginzburg-Landau equation~\cite{aranson2002world}
\bee
 iA_z+(\alpha_1+i\alpha_2)A_{xx}+i\gamma A+(\beta_1+i\beta_2)|A|^2A=0,
\ene
where $A$ is a complex field, $\alpha_{1,2}$, $\beta_{1,2}$, and $\gamma$ are real parameters, is one of the most universal and significant nonlinear wave models in many areas of the physics community, describing all kinds of nonlinear phenomena, such as superfluidity, superconductivity, hydrodynamics, plasmas, reaction-diffusion systems, quantum field theory and Bose-Einstein condensation (BEC), liquid crystals, and strings in the field theory and other physical contexts~\cite{aranson2002world,ipsen2000amplitude,van2003coherent}. The CGL equation can be regarded as a dissipative extension of the conservative nonlinear Schr\"odinger equation describing nonlinear optics, BEC, and waves on deep water. The CGL equation can support stable spatial patterns on account of
the simultaneous balance of gain and loss, as well as nonlinearity versus dispersion or diffraction.
Intriguingly, a vast variety of applications and physical properties in the CGL equations are well elaborated in nonlinear optics~\cite{ferreira2000stable,mandel2004transverse,rosanov2005two,weiss2007pattern,akhmediev2007spatiotemporal,he2012soliton,mihalache2015localized}, where various types of dissipative solitons emerge and are analyzed in detail, including multi-peak solitons~\cite{akhmediev1997multisoliton}, exploding
solitons~\cite{akhmediev2003exploding,soto2005exploding}, pulsating solitons~\cite{tsoy2005bifurcations}, chaotic solitons~\cite{akhmediev2001pulsating}, two-dimensional vortical solitons~\cite{skarka2010varieties}, three-dimensional spatiotemporal optical
solitons~\cite{mihalache2005stable,mihalache2007stability,akhmediev2007spatiotemporal}, accessible solitons~\cite{he2013accessible}, and
lattice solitons~\cite{he2013lattice,he2014localized}.

Meanwhile, we have to mention that, the $\PT$-symmetry~\cite{bender1998real,bender2003must,bender2007making}, put forward by Bender and coworkers in 1998, is a extremely crucial property and widely applied to the complex potentials to support all-real linear spectra ~\cite{bender1998real,ahmed2001real} or stable nonlinear localized modes~\cite{musslimani2008optical,yan2015spatial,yan2013complex,wen2015dynamical,yan2015solitons,lumer2013nonlinearly,
nixon2012stability,achilleos2012dark,shi2011bright}. Many fascinating features and properties related to $\PT$ behaviors such as the celebrated
$\PT$-symmetry breaking phenomenon have been observed or demonstrated in optical experiments~\cite{guo2009observation,rueter2010observation,regensburger2012parity,castaldi2013pt,regensburger2013observation,
peng2014parity,zyablovsky2014pt,chen2016pt,takata2017pt}. Indeed, the $\PT$-symmetric structure can be easily achieved in optics by including a combination of the optical gain and loss regions in the refractive-index guiding geometry~\cite{ultanir2004dissipative,musslimani2008optical}.
 Particularly, in the periodic optical lattice potentials, a great number of novel $\PT$-symmetric behaviors have also been experimentally observed such as the double refraction, secondary emissions, power oscillation, and phase singularities~\cite{makris2008beam,makris2011pt,makris2010pt}.
In the last few years, a great deal of attention has been concentrated on exploring the one- and multi-dimensional solitons and stability in all stripes of optical potentials, including the harmonic potential~\cite{zezyulin2012nonlinear}, Scarf-II potential~\cite{musslimani2008optical,musslimani2008analytical,yan2013complex,yan2015spatial,dai2014stable}, Rosen-Morse potential~\cite{midya2013nonlinear}, Gaussian potential~\cite{hu2011solitons,achilleos2012dark,yang2014symmetry}, super-Gaussian potential~\cite{jisha2014influence},
optical lattices or
super lattices~\cite{abdullaev2011solitons,nixon2012stability,moiseyev2011crossing,lumer2013nonlinearly,jisha2014nonlocal,wang2016two}, photonic
systems~\cite{suchkov2016nonlinear},
time-dependent harmonic-Gaussian potential~\cite{yan2015solitons}, sextic anharmonic double-well potential~\cite{wen2015dynamical},
the double-delta potential~\cite{cartarius2012model,single2014coupling}, and
etc.~\cite{burlak2013stability,bludov2013stable,fortanier2014dipolar,dizdarevic2015cusp,dai2017localized}. Recently, $\mathcal{PT}$-symmetric stable nonlinear localized modes and dynamics were also elucidated in the generalized Gross-Pitaevskii (GP) equation with a variable group-velocity coefficient~\cite{yan2016stable}, the third-order nonlinear Schr\"odinger equation (NLSE)~\cite{chen2016solitonic}, the NLSE with position-dependent effective masses~\cite{chen2017families}, the derivative NLSE~\cite{chen2017stable}, the NLSE with generalized nonlinearities~\cite{yan2017nonlinear}, the nonlocal NLSE~\cite{wen2017solitons}, and the NLSE with spatially-periodic momentum modulation~\cite{chen2018one}.

Besides, solitons and their stability in the NLSE with lots of non-$\PT$-symmetric complex potentials have been investigated theoretically ~\cite{tsoy2014stable,konotop2014families,nixon2016bifurcation,kominis2015soliton,kominis2015dynamic,yang2016stability,konotop2016nonlinear,yan2016stable}.
Notice that rich dynamics of spatial dissipative solitons in the cubic-quintic CGL equation with $\PT$-symmetric periodic potential have been discussed too~\cite{he2013lattice,he2014localized}. However, the non-$\PT$-symmetric potentials in the CGL equation have scarcely been studied.
Therefore, in this paper we aim to demonstrate that a broad class of $\PT$-symmetric stable exact solitons can exist in the cubic CGL model with the non-$\PT$-symmetric potentials. We also find that the non-$\PT$-symmetric potential can be bifurcated out from the $\PT$-symmetric potential by regulating the related potential parameters, which thus is called the near $\PT$-symmetric potential. Furthermore, in the context of CGL model, various dynamical properties associated with the exact solitons are also analyzed and elucidated in detail under the near $\PT$-symmetric potential.
These results are beneficial for applying them in the related experimental designs.

 \begin{figure}[!t]
 	\begin{center}
 	\hspace{-0.05in}{\scalebox{0.6}[0.6]{\includegraphics{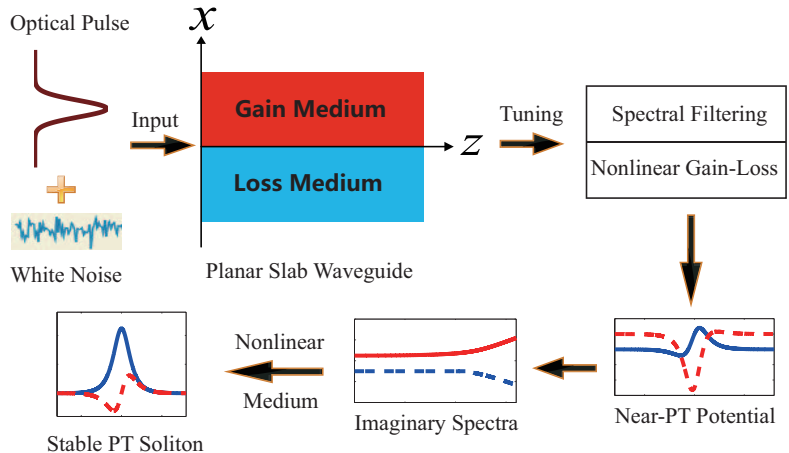}}}
 	\end{center}
 	\vspace{-0.15in} \caption{\small (color online). Schematic of simple experimental design apparatus described by
the complex CGL equation with gain-and-loss distribution. }
  \label{expd}
 \end{figure}

\section{$\PT$-symmetric Physical model}

When an optical pulse with white noise goes through the planar slab waveguide, the upper part will suffer energy gain while the lower part experience energy loss (see Fig.~\ref{expd}). We predicted the unstable propagation of the pulse, if the gain and loss are unbalanced (which can be realized by regulating the spectral filtering parameter and nonlinear gain-loss coefficient), because the linear system with a complex diffraction coefficient always has no purely-real spectrum. However, such a system can support a wide range of stable PT-symmetric solitons in the complex-coefficient Kerr medium, even though the complex refractive index distribution is non-PT-symmetric. To theoretically verify the idea, we begins by considering the spatial beam transmission in an cubic-nonlinear optical medium described by the following CGL equation with complex
potentials~\cite{he2012soliton,he2013lattice}
\bee\label{nls}
 iA_z+(\alpha_1+i\alpha_2)A_{xx}+[V(x)+iW(x)]A+(\beta_1+i\beta_2)|A|^2A=0,
\ene
where $A\equiv A(x,z)$ is the normalized envelope of the complex light field, $z$ denotes the propagation distance, and $x$ represents the scaled spatial coordinate; for the convenience of study, both the diffraction coefficient $\alpha_1$ and Kerr-nonlinearity coefficient $\beta_1$ are fixed with $\alpha_1=\beta_1=1$ in the paper; the real parameter $\alpha_2$ can be used to describe the spectral filtering or linear parabolic gain ($\alpha_2>0$), and the real constant $\beta_2$ accounts for the nonlinear gain/loss processes. Different from the traditional GL
equations~\cite{akhmediev1997multisoliton,akhmediev2003exploding,soto2005exploding,tsoy2005bifurcations,akhmediev2001pulsating,soto2001simultaneous}, we introduce the complex potential $V(x)+iW(x)$ instead of the constant linear gain-loss coefficient. Compared with those discussed in Refs.~\cite{he2012soliton,he2013lattice}, the spectral filtering coefficient $\alpha_2$ is added such that it is possible to exhibit some distinct behaviors. The complex potential $V(x)+iW(x)$ is $\PT$-symmetric provided that $V(x)=V(-x)$ and $W(-x)=-W(x)$. Physically, the real-valued external potential $V(x)$ is closely related to the refractive index waveguide while $W(x)$ characterizes the amplification (gain) or absorption (loss) of light beam in the optical material.  On account of the occurrence of complex coefficients, Eq.~(\ref{nls}) is not invariant any more under the action of $\PT$ operator, with the operators $\mathcal{P}$
and $\mathcal{T}$ respectively defined by $\mathcal{P}: x\rightarrow -x; \mathcal{T}: i\rightarrow -i, z\rightarrow -z$. Besides, Eq.~(\ref{nls}) can also be rewritten as another variational form $iA_z=\delta\mathcal{H}(A)/\delta A^*$, where the Hamiltonian
$\mathcal{H}(A)=\int_{-\infty}^{+\infty}\left\{(\alpha_1+i\alpha_2)|A_x|^2 -[V(x)+i W(x)]|A|^2-\frac{1}2(\beta_1+i\beta_2)|A|^4\right\}dx$ and the asterisk denotes the complex conjugate. If we define the optical power of Eq.~(\ref{nls}) as $P(z)=\int_{-\infty}^{+\infty}|A(x,z)|^2dx$, then one can elicit immediately that the power evolves by $P_z=\int_{-\infty}^{+\infty}[2\alpha_2|A_x|^2-\alpha_2(|A|^2)_{xx}- 2W(x)|A|^2-2\beta_2|A|^4]dx$.
Moreover, when setting $z\rightarrow z$ (cavity round-trip number) and $x\rightarrow t$ (retarded time) in Eq.~(\ref{nls}), the aforementioned model may be used to describe the passively mode-locked lasers too~\cite{akhmediev2005dissipative}.

\section{Theoretical analysis}

\subsection{Stationary solitons and linear-stability theory}

 Stationary soliton solutions are explored in the form $A(x,z)=\phi(x)e^{iqz}$, where $q$ is a real propagation constant. Plugging it into Eq.~(\ref{nls}), one can derive at once that the complex localized field-amplitude function $\phi(x)$ ($\lim_{|x|\to\infty} \phi(x)=0$ for $\phi(x)\in C[x]$) satisfies the following second-order ordinary differential equation (ODE) with complex coefficients
\bee\label{stat}
\left[(\alpha_1+i\alpha_2)\frac{d^2}{dx^2}+V(x)+iW(x)+(\beta_1+i\beta_2)|\phi|^2\right]\phi=q\phi,
\ene
In general, exact nonlinear localized modes of Eq.~(\ref{stat}) can be attainable only for some certain combinations of the values of the parameters~\cite{akhmediev1995novel,akhmediev1996singularities}. Therefore, some useful numerical techniques are necessary to find its stationary soliton solutions~\cite{soto2001simultaneous,yang2010nonlinear}.

We readily know that every stationary nonlinear mode is a singular point of the nonlinear dynamical system in an infinite-dimensional phase space. To investigate the linear stability of stationary soliton, we perturb the solution in the vicinity of the singular point
\bee\label{pert}
A(x,z)=\{\phi(x)+\epsilon [f(x)e^{\delta z}+g^*(x)e^{\delta^* z}]\}e^{iq z},
\ene
where $|\epsilon|\ll 1$, $f(x)$ and $g(x)$ are the perturbation eigenfunctions, and $\delta$ reveals the perturbation growth rate. Inserting this perturbed solution (\ref{pert}) into Eq.~(\ref{nls}) and linearizing with respect to $\epsilon$, we obtain the following linear-stability eigenvalue problem
\bee \label{sta}
i\left[\begin{array}{cc}   \hat{L}_1 & \hat{L}_2 \vspace{0.05in}\\   -\hat{L}_2^* & -\hat{L}_1^* \\  \end{array}\right]
\left[  \begin{array}{c}    f(x) \vspace{0.05in} \\    g(x) \\  \end{array} \right]
=\delta \left[  \begin{array}{c}   f(x) \vspace{0.05in}\\    g(x) \\  \end{array}\right],
\label{stable}
\ene
where $\hat{L}_1=(\alpha_1+i\alpha_2)\partial_{xx}+V(x)+iW(x)+2(\beta_1+i\beta_2)|\phi|^2-q$ and $\hat{L}_2=(\beta_1+i\beta_2)\phi^2$. It is more than evident that the nonlinear localized modes are linearly unstable if $\delta$ possesses a positive real part, otherwise they are linearly
stable. In practice, the linear stability is determined by the maximal value of real parts of the linearized eigenvalues $\delta$, i.e. $\max[\Re(\delta)]$. The full stability spectrum of $\delta$ can be numerically computed by the Fourier collocation method (see~\cite{yang2010nonlinear}).

\subsection{Near $\PT$-symmetric Scarf-II potential}

In what follows, we initiate our analysis by introducing the following near $\PT$-symmetric Scarf-II potential in this form
\bee\label{rp}
 \begin{array}{l}
 V(x)=V_0\sech^2(x)-\frac{\alpha_2}{\alpha_1}\, W_0\sech(x)\tanh(x), \vspace{0.1in} \\
  W(x)=W_0\sech(x)\tanh(x)+W_1\sech^2(x)-\alpha_2,
  \end{array}
\ene
where $W_1=\left(\alpha_2-\alpha_1\frac{\beta_2}{\beta_1}\right)\left(2+\frac{W_0^2}{9\alpha_1^2}\right)+\frac{\beta_2}{\beta_1} V_0$, the real parameters both $V_0$ and $W_0$ can be used to modulate the strength of the real and imaginary parts of the complex potential. It is evident that the aforementioned complex potential $V(x)+iW(x)$ reduces to the usual $\PT$-symmetric Scarf-II potential at once if $\alpha_2=\beta_2=0$, meanwhile
Eq.~(\ref{nls}) becomes the well-known $\PT$-symmetric nonlinear Schr\"odinger equation. However, when $\alpha_2$ or $\beta_2$ is perturbed around
the origin in the $(\alpha_2, \beta_2)$ space, Eq.~(\ref{nls}) turns into the complex cubic GL equation and the corresponding complex potential is not
$\PT$-symmetric any more. We call such a complex potential is near $\PT$-symmetric in the $(\alpha_2, \beta_2)$ parameter space, because $V(x)+iW(x)$
tends to be $\PT$-symmetric as $(\alpha_2, \beta_2)\rightarrow (0, 0)$. In addition, it is also apparent that the aforementioned complex potential
possesses even symmetry if $W_0=0$, due to $V(x)=V(-x)$ and $W(x)=W(-x)$.

\section{NUMERICAL SIMULATIONs}

\subsection{Unbroken or broken near $\PT$-symmetric phases}

\begin{figure}[!t]
 	\begin{center}
 	\vspace{0.05in}
 	\hspace{-0.05in}{\scalebox{0.8}[0.8]{\includegraphics{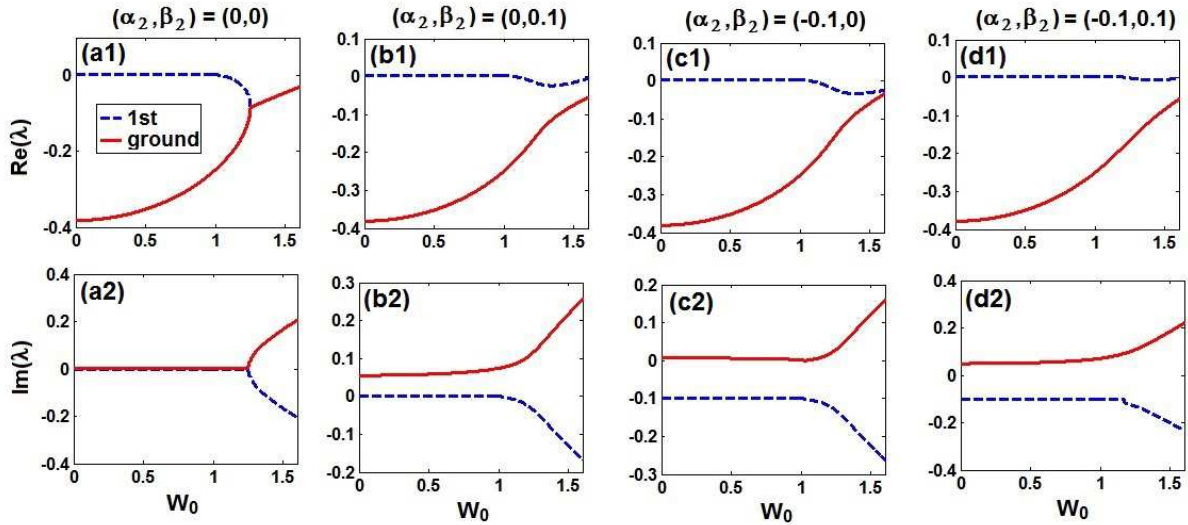}}}
 	\end{center}
 	\vspace{-0.25in} \caption{\small (color online). Real and imaginary components of the first two lowest energy eigenvalues $\lambda$ of the linear spectral problem (\ref{ls}) as a function of $W_0$ at $V_0=1$: (a1, a2) $(\alpha_2, \beta_2)=(0, 0)$, (b1, b2) $(\alpha_2, \beta_2)=(0, 0.1)$, (c1, c2) $(\alpha_2, \beta_2)=(-0.1, 0)$, (d1, d2) $(\alpha_2, \beta_2)=(-0.1, 0.1)$, in the near $\PT$-symmetric potential (\ref{rp}).}
  \label{spe}
 \end{figure}

Next we turn to investigate the unbroken or broken phases in the near $\PT$-symmetric potential (\ref{rp}) by considering the following linear eigenvalue problem
\bee \label{ls}
 L\Phi(x)=\lambda\Phi(x),\qquad L=-(\alpha_1+i\alpha_2)\frac{d^2}{dx^2}\!+\! V(x)\!+\! iW(x),
\ene
where $\lambda$ and $\Phi(x)$ stand for the eigenvalue and eigenfunction, respectively. Unluckily, abundant numerical results indicate that
unbroken-phase regions barely exist in the potential parameter $(V_0, W_0)$ space, unless $(\alpha_2, \beta_2)=(0, 0)$ which means the linear operator $L$ is $\PT$-symmetric. It fully reveals that the $\PT$ symmetry for a complex potential is of great importance to ensure the real property of the
corresponding eigenvalue spectrum. For illustration, we take $V_0=1$ in Eq.~(\ref{rp}) to illustrate the spontaneous symmetry-breaking process,
which stems from the collision of the first few lowest energy levels. Fig.~\ref{spe}(a1, a2) display the classical situation of $\PT$-symmetric Scarf-II potential, with the phase-transition point $W_0=1.25$. However, only if $\alpha_2$ or $\beta_2$ is not zero, there always exist at least an imaginary eigenvalue in the linear spectra (see the last three columns of Fig.~\ref{spe}). It is easy to observe that the absolute value of the
imaginary part of these complex eigenvalues tends to increase monotonically as $W_0$ grows. Hence a useful conclusion can be reached that
nonzero $\alpha_2$, $\beta_2$, and large values of $|W_0|$ are all extremely adverse to the generation of a full-real spectrum,
which leads to the breaking of phases.

 \begin{figure*}[!t]
 	\begin{center}
 	\vspace{0.05in}
 	\hspace{-0.05in}{\scalebox{0.8}[0.8]{\includegraphics{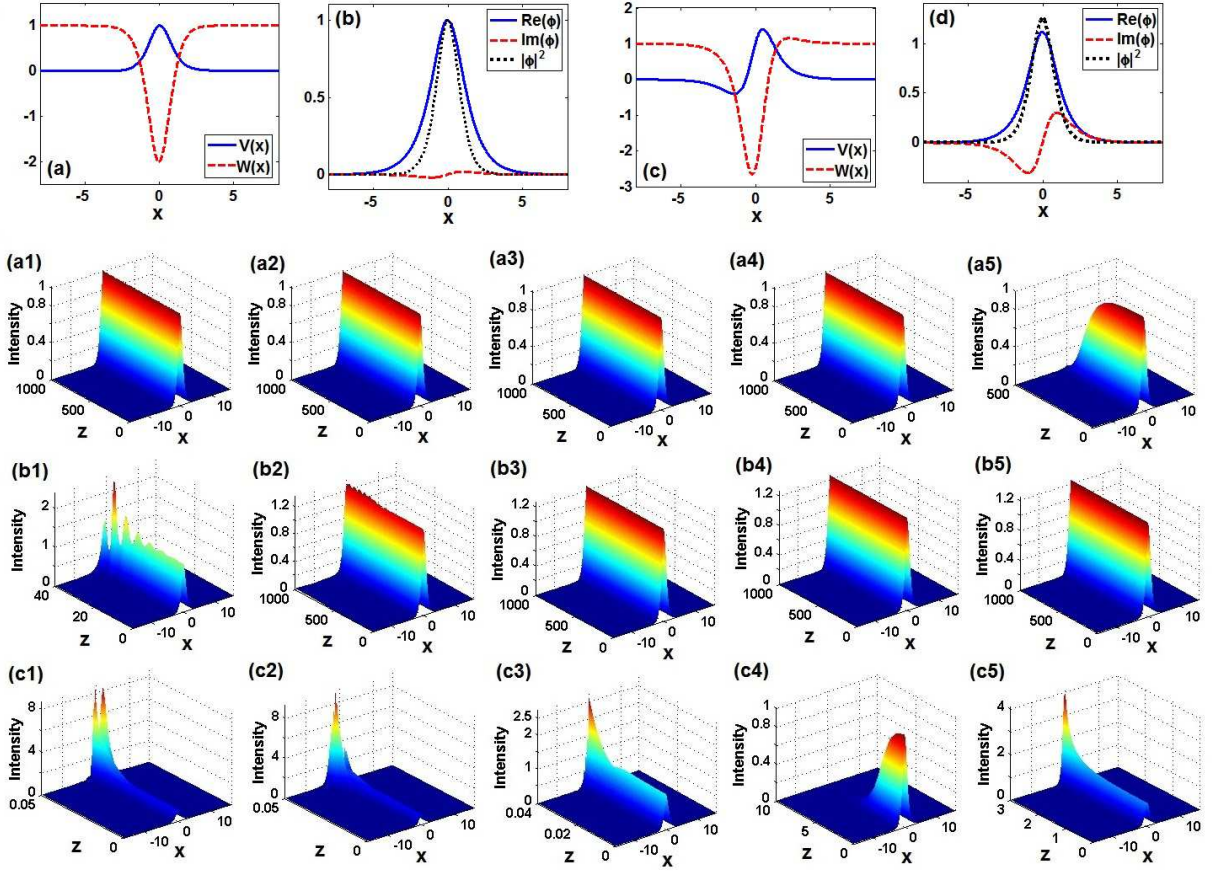}}}
 	\end{center}
 	\vspace{-0.3in} \caption{\small (color online). Profiles of the near $\PT$-symmetric potential (\ref{rp}) and the corresponding soliton solutions with $\alpha_2=-1, \beta_2=1$: (a, b) $W_0=0.1$, (c, d) $W_0=1.5$. Evolutions of the exact solitons (\ref{nlm}) with $W_0=0.1$ in the second row while $W_0=1.5$ in the third row: (a1, b1) $(\alpha_2, \beta_2)=(0, 0)$, (a2, b2) $(\alpha_2, \beta_2)=(0, 1)$, (a3, b3) $(\alpha_2, \beta_2)=(-1, 0)$, (a4, b4) $(\alpha_2, \beta_2)=(-1, 1)$, (a5, b5) $(\alpha_2, \beta_2)=(-0.2, -0.01)$. Unstable evolutions with $W_0=0.1$ in the last row: (c1) $(\alpha_2, \beta_2)=(1, 1)$, (c2) $(\alpha_2, \beta_2)=(1, 0)$, (c3) $(\alpha_2, \beta_2)=(1, -1)$, (c4) $(\alpha_2, \beta_2)=(0, -1)$, (c5) $(\alpha_2, \beta_2)=(-1, -1)$.}
 	\label{exa}
 \end{figure*}

 \begin{figure*}[!t]
 	\begin{center}
 	\vspace{0.05in}
 	\hspace{-0.05in}{\scalebox{0.8}[0.8]{\includegraphics{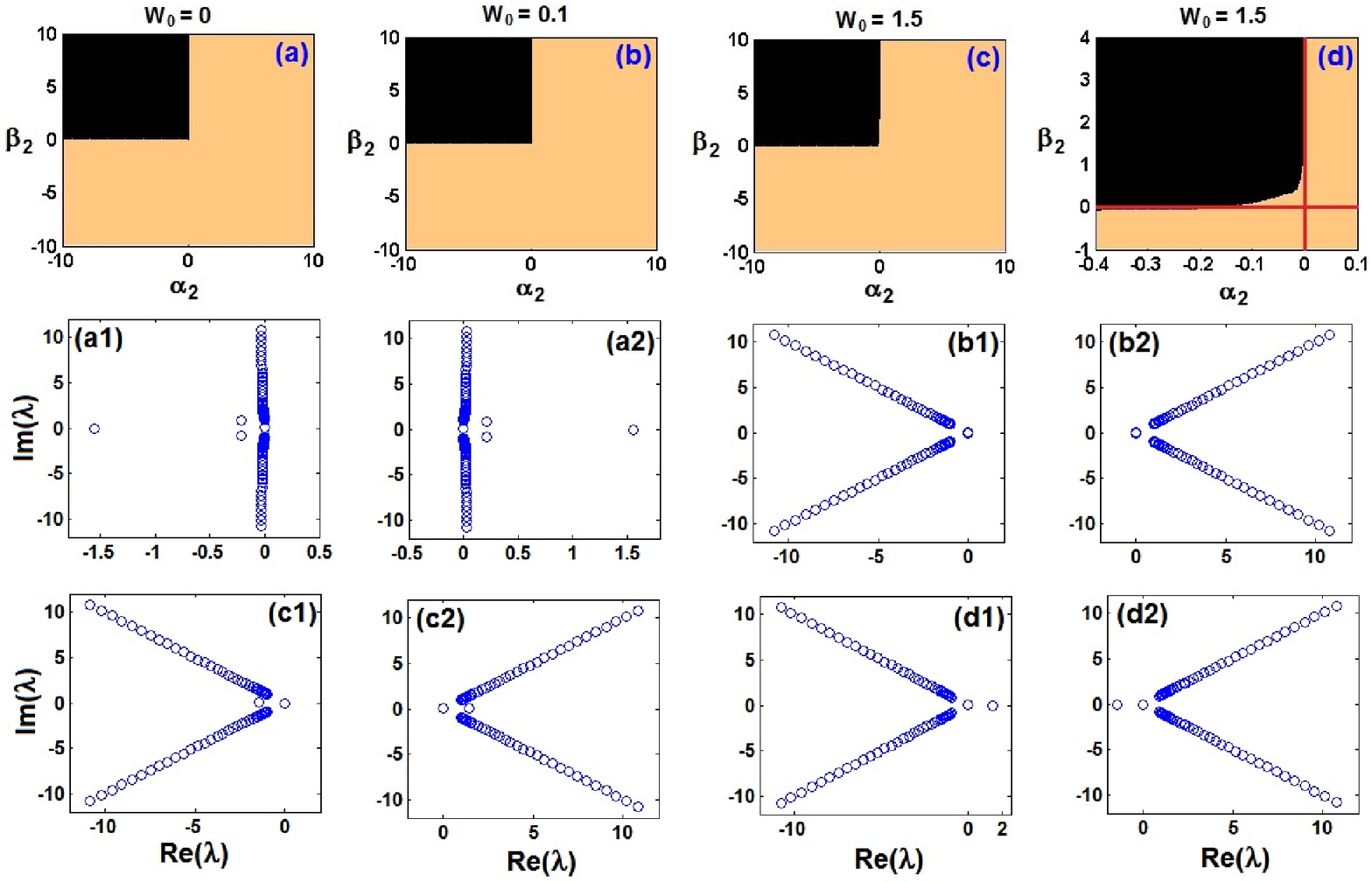}}}
 	\end{center}
 	\vspace{-0.2in} \caption{\small (color online). Linear-stability maps [cf. Eq.~(\ref{sta})] of the exact solitons (\ref{nlm}) in the $(\alpha_2, \beta_2)$ space [only black and dark regions denote stable solitons]: (a) $W_0=0$, (b) $W_0=0.1$, (c, d) $W_0=1.5$, where (d) clearly indicates the concrete linear-stability situation around the original point in (c). Linear-stability spectra with $W_0=0.1$: (a1) $(\alpha_2, \beta_2)=(0, 1)$,  (a2) $(\alpha_2, \beta_2)=(0, -1)$, (b1) $(\alpha_2, \beta_2)=(-1, 0)$, (b2) $(\alpha_2, \beta_2)=(1, 0)$, (c1) $(\alpha_2, \beta_2)=(-1, 1)$, (c2) $(\alpha_2, \beta_2)=(1, -1)$, (d1) $(\alpha_2, \beta_2)=(-1, -1)$, (d2) $(\alpha_2, \beta_2)=(1, 1)$.}
 	\label{stab}
 \end{figure*}

\subsection{Analytical solitons and dynamical stability}

In the current section, we turn to discuss the stationary soliton solutions of Eq.~(\ref{stat}) under the near $\PT$-symmetric potential (\ref{rp}). Similar to the analytical theory in the nonlinear Schr\"odinger equation~\cite{musslimani2008analytical,musslimani2008optical}, the exact nonlinear
localized mode of Eq.~(\ref{stat}) corresponding to the propagation constant $q=\alpha_1$ can be obtained in the following form
\bee\label{nlm}
\phi(x)= \sqrt{(\alpha_1[2+W_0^2/(9\alpha_1^2)]-V_0)/\beta_1}\, \sech(x) e^{i \frac{W_0}{3\alpha_1}\arctan[\sinh(x)]},
\ene
It is noteworthy that the exact soliton solution above always keeps invariant, no matter how $\alpha_2$ and $\beta_2$ change in the potential (\ref{rp}).
However, the variation of $\alpha_2$ and $\beta_2$ can dramatically change the stability of the soliton solution (\ref{nlm}), which will be demonstrated in the following. When $\alpha_1$ and $\beta_1$ are fixed, we can regulate the potential parameters $V_0$ and $W_0$ to control the profiles of the complex potential (\ref{rp}) and soliton solution (\ref{nlm}). For convenience, we always fix $V_0=1$ in the following discussion. When we choose a smaller value of $W_0=0.1$, the potential (\ref{rp}) looks almost even symmetric (see Fig.~\ref{exa}(a)); if we further increase $W_0$ to $1.5$, the asymmetric phenomenon of the potential (\ref{rp}) begins to become obvious (see Fig.~\ref{exa}(c)). Nonetheless, the corresponding two solitons are $\PT$-symmetric, which are exhibited in Fig.~\ref{exa}(b, d), which indicates that at this moment, just the eigenstate of the system no longer meet the PT symmetry, the system still shows the characteristics of the conserved system. One of the possible physical explanations we believe is that in the case of self-focusing nonlinearities in the system, the increase of the nonlinear refractive index caused by the self-focusing effect and the real part of the linear potential function work together, resulting in the soliton is PT symmetric even the value of $W_0$ above the phase-transition point.

In order to explore the stability of the soliton (\ref{nlm}), direct beam propagation method is used and we take the soliton (\ref{nlm}) with some $2\%$
white noise as the initial condition to simulate the wave transmission. First, we show that the soliton in Fig.~\ref{exa}(b) is stable while that in Fig.~\ref{exa}(d) is unstable as $(\alpha_2, \beta_2)=(0, 0)$ (see Fig.~\ref{exa}(a1, b1)). A important reason is that the former lies in the parameter regions with the unbroken $\PT$-symmetric phase, whereas the latter with the broken $\PT$-symmetric phase. Second, increasing $\beta_2$ to
positive values or decreasing $\alpha$ to negative values is more favorable to the stability of the soliton (see Fig.~\ref{exa}(a2-a4, b2-b4)). Third, at some exceptional points in the $(\alpha_2, \beta_2)$ space, the growth of $W_0$ can also increase soliton stability, which is a novel phenomenon and breaks the traditional mindset (compare Fig.~\ref{exa}(a5) with (b5)). Moreover, we test out that for small values of $W_0$, the soliton (\ref{nlm}) is usually stable in the second quadrant of the $(\alpha_2, \beta_2)$ space (including the nonnegative vertical axis and nonpositive
horizontal axis), beyond which the soliton immediately becomes extremely unstable (see Fig.~\ref{exa}(c1-c5)). More importantly, these nonlinear-propagation stability results can be predicted and validated by the forthcoming linear stability analysis.

\subsection{Linear stability and spectral property}

According to the above-mentioned linear-stability theory, we investigate that the influence of $W_0$ on soliton stability in the whole
$(\alpha_2, \beta_2)$ space. We can observe apparently from Fig.~\ref{stab}(a, b) that when $W_0$ is small to some extent, the stable domains of
the soliton (\ref{nlm}) are all located in the second quadrant of the $(\alpha_2, \beta_2)$ space (including the corresponding axes), which may be why one usually assumes $\alpha_2<0$ and $\beta_2>0$ in the study of complex GL equation. As $W_0$ rises, most stable areas still remain in the second
quadrant (see Fig.~\ref{stab}(c)); meanwhile, the unstable regions also begin to emerge in the vicinity of the origin, which can be observed more clearly in Fig.~\ref{stab}(d). Noting that at the origin point $(\alpha_2, \beta_2)=(0, 0)$, the soliton is unstable though the corresponding potential is $\PT$-symmetric. However, we can regulate the parameter $\alpha_2$ or $\beta_2$ to make the soliton keep stable, although at this moment
the potential may not satisfy $\PT$ symmetry. In addition, Fig.~\ref{stab}(d) also exhibits that, below and near the negative horizontal axis,
stable solitons can be found too, as has been shown in Fig.~\ref{exa}(b5). This is possible because the beam can change the refractive index profile through optical nonlinearity and further adjust the amplitude to maintain the stable transmission.

Another intriguing phenomenon is closely related to the concrete linear-stability spectrum. It is well-known that if $(\alpha_2, \beta_2)=(0, 0)$ (which means the corresponding potential is $\PT$-symmetric), the linear-stability spectrum is generally symmetric with respect to the real and imaginary
axes, with the final (or tail) eigenvalues distributed on the imaginary axis. However, the positive (negative) values of $\beta_2$ can generate several or finite pairs of complex-conjugate eigenvalues on the left (right) side of the imaginary axis, as is shown in Fig.~\ref{exa}(a1, a2).
In contrast, the negative (positive) values of $\alpha_2$ can lead to infinite pairs of complex-conjugate eigenvalues on the left (right) side of the
imaginary axis (see Fig.~\ref{exa}(b1, b2)). The combined-action effect of $\alpha_2$ and $\beta_2$ has also been displayed in
Fig.~\ref{exa}(c1, c2, d1, d2). In brief, the linear-stability spectrum is only symmetric with respect to the real axis, if $\alpha_2$ or $\beta_2$ is nonzero; only the nonnegative $\beta_2$ and and nonpositive $\alpha_2$ make the real parts of the spectrum admit the nonpositive maximum value, which
contributes to the generation of a stable soliton (see Fig.~\ref{exa}(a1, c1)); more importantly, through lots of numerical tests, one can summarize
that the linear-stability spectrum at $(\alpha_2, \beta_2)$ and that at $(-\alpha_2, -\beta_2)$ are symmetric with regard to the imaginary axis,
that is, the centrosymmetric two points in the $(\alpha_2, \beta_2)$ parameter space enjoy imaginary-axis symmetric (or even-symmetric)
linear-stability spectra.

 \begin{figure}[!t]
 	\begin{center}
 	\vspace{0.05in}
 	\hspace{-0.05in}{\scalebox{0.6}[0.6]{\includegraphics{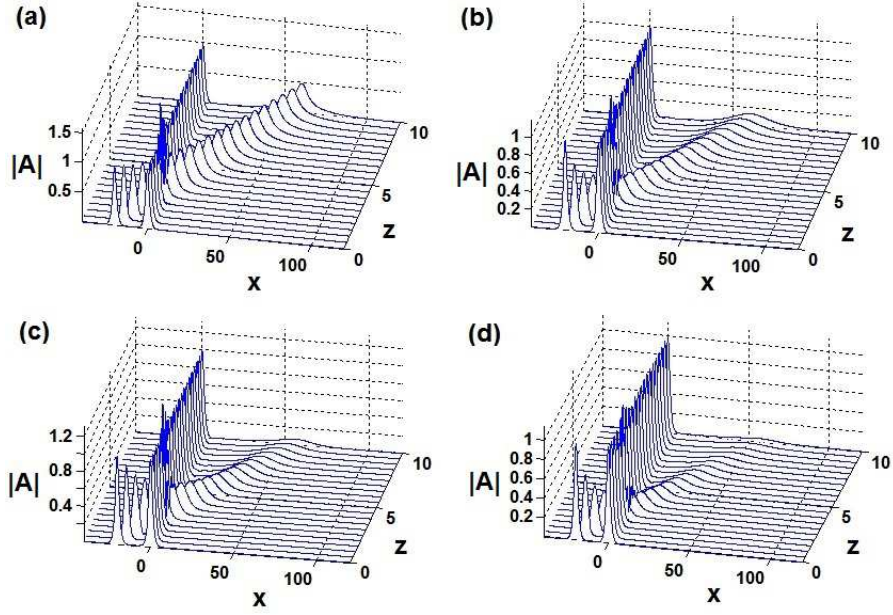}}}
 	\end{center}
 	\vspace{-0.15in} \caption{\small (color online). Collisions between the exact bright soliton (\ref{nlm}) and the boosted sech-shaped or rational solitary pulse, produced by the simulation of Eq.~(\ref{nls}), with the initial input $A(x,0)=\phi(x)+ {\rm sech}(x+20) \, e^{4ix}$: (a) $(\alpha_2, \beta_2)=(0, 0)$, (b) $(\alpha_2, \beta_2)=(0, 1)$, (c) $(\alpha_2, \beta_2)=(-0.01, 0)$, (d) $(\alpha_2, \beta_2)=(-0.01, 1)$. Here $\phi(x)$ is given by Eq.~(\ref{nlm}) in the potential parameters $V_0=1, W_0=0.1$. }
 	\label{col}
 \end{figure}

 \begin{figure}[!t]
 	\begin{center}
 	\vspace{0.05in}
 	\hspace{-0.05in}{\scalebox{0.5}[0.5]{\includegraphics{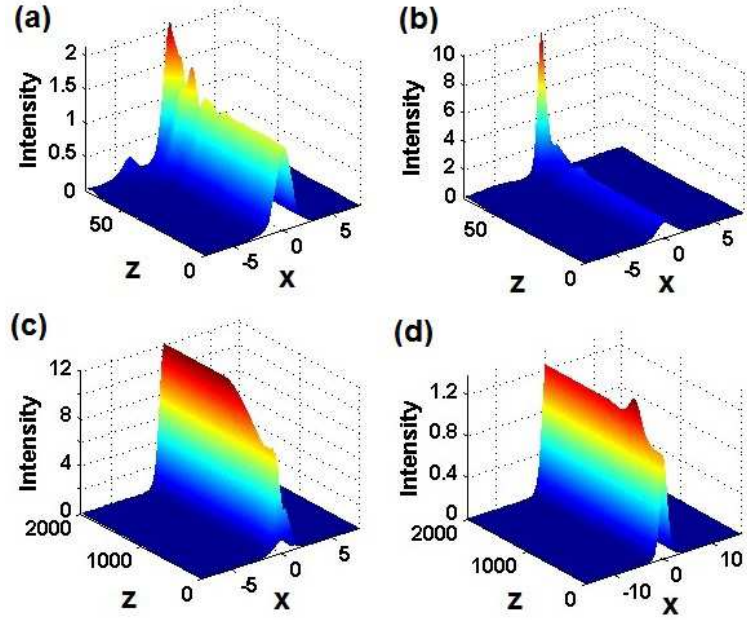}}}
 	\end{center}
 	\vspace{-0.15in} \caption{\small (color online). Excitations of exact nonlinear localized modes [cf. Eq.~(\ref{tnls})]:
    (a) $\alpha_{21}= 0, \alpha_{22}= -1, \beta_2= 0$, (b) $\alpha_2=0, \beta_{21}= 0, \beta_{22}= 1$, (c) $\alpha_{21}= 0, \alpha_{22}= -1, \beta_{21}= 0, \beta_{22}= 1$, other parameter is $W_0= 1.5$;
    (d) $W_{01}= 0.1, W_{02}= 1.5$, other parameters are $\alpha_2= -0.2, \beta_2= -0.01$.
  }
 	\label{exc}
 \end{figure}

\subsection{Influence of exotic solitary wave on the stable exact soliton}

To further examine the robustness of the exact nonlinear localized modes (\ref{nlm}), we explore their interactions with boosted sech-shaped  solitary pulses. Without loss of generality, we assume the exotic solitary wave is always in the form ${\rm sech}(x+20) \, e^{4ix}$. For illustration, we set $V_{0}=1, W_{0}=0.1$ and first choose the exact bright soliton (\ref{nlm}) with the parameter $(\alpha_2, \beta_2)=(0, 0)$ and take the following initial condition
$A(x,0)=\phi(x)+{\rm sech}(x+20) \, e^{4ix}$ to simulate the wave propagation governed by Eq.~(\ref{nls}). The result of interaction reveals that
the exact bright soliton can remain stable without any change of shape before and after collision, only with mild dissipation of the exotic wave, as is displayed in Fig.~\ref{col}(a). When we increase $\beta_2$ or decrease $\alpha_2$ a little, the shape of the exact soliton still
doesn't change at all, whereas the amplitude of the exotic solitary wave declines rapidly (see Fig.~\ref{col}(b, c)). The combined action of increasing
$\beta_2$ and decreasing $\alpha_2$ only aggravates the rapid-decline process of the amplitude of the exotic solitary wave while has no influence on
the stable propagation of the exact soliton (see Fig.~\ref{col}(d)). That can be explained by considering the relationship between the coefficients $\alpha_2$ and $\beta_2$ when the $W_{0}$ below the phase-transition point. The nonlinear gain/loss of the exact soliton is greater than that of the linear parabolic gain, so the exotic solitary wave is continuously diffused in the transmission process, and the lager difference between the two parameters is, the more serious the diffusion is.

\subsection{Excitations of the exact soliton}

In the present section, we turn to elaborate the excitations of the exact bright soliton (\ref{nlm}) by making the parameters rely on the
propagation distance: $\alpha_2\rightarrow \alpha_2(z)$ or $\beta_2\rightarrow \beta_2(z)$ (cf. Refs.~\cite{yan2015solitons,chen2017stable}).
It requires that the simultaneous adiabatic switching is imposed on the near-$\PT$-symmetric potential (\ref{rp}) and complex coefficients of Eq.~(\ref{nls}), regulated by
\bee\label{tnls}
 iA_2+[\alpha_1+i\alpha_2(z)]A_{xx}+[V(x,z))+iW(x,z)]A+[\beta_1+i\beta_2(z)]|A|^2A=0,
\ene
where $V(x,z),\, W(x,z)$ are given respectively by Eqs.~(\ref{rp}) with $\alpha_2\rightarrow \alpha_2(z)$ and $\beta_2\rightarrow \beta_2(z)$.
For convenience, both $\alpha_2(z)$ and $\beta_2(z)$ are selected as the following unified form
\be\label{excite-s}\epsilon(z)=
\begin{cases}
\dfrac{(\epsilon_2-\epsilon_1)}2 \left[1-\cos\left(\dfrac{\pi z}{1000}\right)\right]+\epsilon_1, & \text{$0\leq z<1000$},\\
\epsilon_2, & \text{$z\geq 1000$}
\end{cases}
\ee
where $\epsilon_{1,2}$ respectively represent the real initial-state and final-state parameters. One can easily examine that the
soliton (\ref{nlm}) with $\alpha_2\rightarrow \alpha_2(z)$ or $\beta_2\rightarrow \beta_2(z)$ don't satisfy Eq.~(\ref{tnls}) any longer,
nevertheless the bright soliton (\ref{nlm}) do solve Eq.~(\ref{tnls}) for both the initial state $z=0$ and excited states $z\geq 1000$.

We first execute a single-parameter excitation of the soliton $A(x,z)$ controlled by Eq.~(\ref{tnls}) via the initial condition determined by Eq.~(\ref{nlm}), with $\beta_2(z)$ given by Eq.~(\ref{excite-s}) and $\alpha_2(z)\equiv \alpha_2$ fixed. Fig.~\ref{exc}(a) displays the excitation or dynamical transformation of the nonlinear mode is unstable due to the unstable initial state, though the final state (\ref{nlm})
is stable in Eq.~(\ref{nls}). The similar situation happens for the excitation of the single-parameter $\alpha_2$ (see Fig.~\ref{exc}(b)). However,
when the two-parameter simultaneous excitation is carried out with both $V_0(z)$ and $W_0(z)$ determined by Eq.~(\ref{excite-s}) concurrently, we can
excite a initially unstable exact nonlinear localized mode given by Eq.~(\ref{nlm}) to another stable exact nonlinear localized mode
as is shown in Fig.~\ref{exc}(c). It can be obviously observed from the amplitude of the intensity that the final stable state in the process of
excitation is not regulated by Eq.~(\ref{nlm}) any more, which is a novel finding.
Moreover, only by modulating $W_0\rightarrow W_0(z)$ determined by Eq.~(\ref{excite-s}), a initially unstable exact nonlinear localized mode given by Eq.~(\ref{nlm}) can also be excited to another stable exact nonlinear localized mode, where the stable final state
satisfies Eq.~(\ref{nlm}) (see Fig.~\ref{exc}(d)).
 \begin{figure}[!t]
 	\begin{center}
 	\vspace{0.05in}
 	\hspace{-0.05in}{\scalebox{0.5}[0.5]{\includegraphics{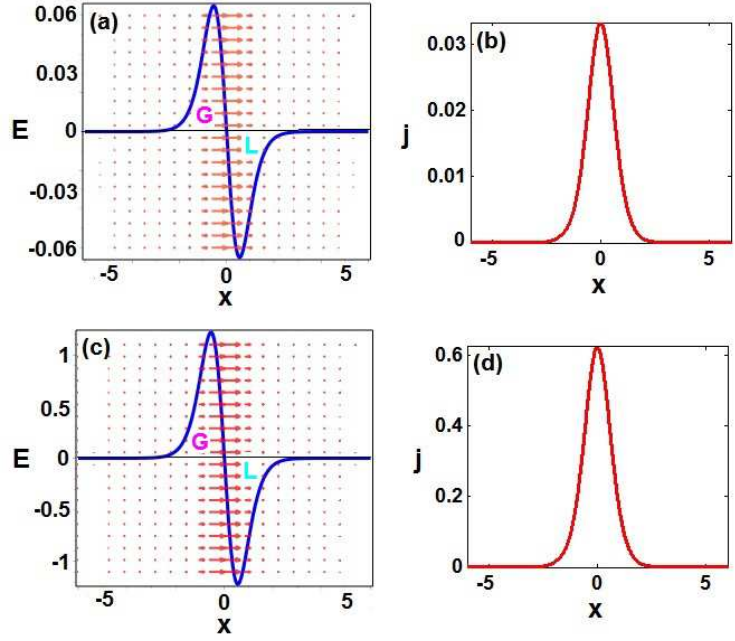}}}
 	\end{center}
 	\vspace{-0.25in} \caption{\small (color online).
     The density of energy generation $E$ and the corresponding energy flux $j$: (a, b) The same parameters are used as Fig.~\ref{exa}(a, b);
     (c, d) the same parameters as Fig.~\ref{exa}(c, d). Here `G' (`L') denotes the gain (loss) region, and the red right
     arrows represent the direction of energy flow from gain to loss regions.}
 	\label{ef}
 \end{figure}

\subsection{Energy flow across the exact soliton}

Last but not least, we also examine the transverse energy flow intensity of the exact soliton determined by Eq.~(\ref{nlm}),
defined by $j(x)=\frac{i}2(AA^*_x-A^*A_x)$. Based on the celebrated continuity relation of the GL equation,
$\frac{\partial\rho}{\partial z} + \frac{\partial j}{\partial x} = E$, where $\rho=|A|^2$ denotes the energy density, we can attain the
density of energy gain or loss
\bee
E=2\alpha_2|A_x|^2-\alpha_2(|A|^2)_{xx}- 2W(x)|A|^2-2\beta_2|A|^4,
\ene
which determines the gain or loss distribution of energy. If these complex coefficients in Eq.~(\ref{nls}) disappear, i.e. $\alpha_2=\beta_2=0$ and
$W(x)\equiv 0$, the system is conservative because of $E=0$, otherwise it is dissipative. The energy of the optical field can be transported laterally from the gain region to the loss region through the effect of phase gradient, so that the whole system maintains the balance of gain and loss effects, which corresponds to a passive system and therefore exhibits Hermitian properties. However, when the eigenvalues enter the complex region, the PT symmetry of the system is broken and the whole gain and loss effects are no longer balanced. The system shows a dissipative effect. For a fixed value of $W_0$ ($W_0=0.1$ without loss of generality), the variation of the parameters $\alpha_2$ and $\beta_2$ basically does not change the gain and loss distribution of energy and the corresponding energy flux, as is displayed in Fig.~\ref{ef}(a, b) by vast numerical tests. However, when $W_0$ rises,
the strength of the gain-loss distribution and the corresponding flux will grow too, but their respective shapes and the flow direction still
remain unchanged, by comparing Fig.~\ref{ef}(c, d) and (a, b). In fact, these findings can be proved by the analytical calculation. For convenience,
we still fix $\alpha_1=\beta_1=V_0=1$, and substitute the exact solution determined by the stationary nonlinear modes (\ref{nlm}) into the aforementioned formulas with respect to $E$ and $j$, then we can obtain $E=-\frac29 W_0(W_0^2+9) \sinh(x)\, \sech^4(x)$ and $j=\frac1{27} W_0(W_0^2+9) \sech^3(x)$, both only related to $W_0$ and independent on $\alpha_2$ and $\beta_2$.
In brief, $\alpha_2$ and $\beta_2$ don't change the generation or loss distribution
of energy and the flux including the magnitude and direction which always flows from gain to loss regions at all;
$W_0$ can regulate their magnitude whereas keep their shapes and the flow direction.

\section{Conclusions and discussions}

In conclusion, we mainly present a class of exact $\PT$-symmetric solitons can reside in the complex Kerr-nonlinear GL equation with a novel category of near-$\PT$-symmetric potentials, where the phase in the linear regime is always symmetry-breaking because of the occurrence of spectral filtering
parameter $\alpha_2$ or nonlinear gain-loss coefficient $\beta_2$. Nonlinear-propagation dynamics and linear-stability analysis reveal that the overwhelming majority of stable solitons are located in the second quadrant of the $(\alpha_2, \beta_2)$ parameter space. A fascinating finding is that the centrosymmetric two points $(\alpha_2, \beta_2)$ and $(-\alpha_2, -\beta_2)$ have imaginary-axis symmetric linear-stability spectra. Moreover, by adiabatically changing $\alpha_2$ and $\beta_2$, we can excite an unstable nonlinear mode to another stable nonlinear mode.
The interactions and energy flow with respect to exact solitons are checked too.

Before closing we would like to mention that in the paper the exact nonlinear localized modes are attained at some special fixed propagation-constant
points. One can further investigate the numerical solitons for other points, including their stability analysis and other significant properties. In
addition, our analysis and methods can also be used to study some more general modes by adding competing nonlinearity, the fourth-order term
$isA_{xxxx}$ ($s$ is associated with higher-order spectral filtering), or other complex potentials into the GL equation, such as the
well-known complex cubic-quintic GL equation and Swift-Hohenberg equation. Finally, it is an open problem that our results presented here may provide the related physical researchers with several helpful theoretical guidance to design relevant experiments in optics or other fields.
\acknowledgments

This work was supported by the NSFC under Grant No.11571346 and CAS Interdisciplinary Innovation Team.

\bibliography{PT-GLE}

\end{document}